\documentclass[aps,prb,twocolumn,superscriptaddress]{revtex4-1}
\usepackage[pdftex]{graphicx}
\usepackage{color,soul}
\usepackage{booktabs}
\usepackage{enumitem}
\usepackage{gensymb}
\usepackage{epstopdf}
\usepackage{bm}
\usepackage{amsmath,amsthm,amssymb}
\usepackage{dcolumn}

\begin{document}

\title{Tantalum-Palladium: hysteresis-free optical hydrogen sensor over 7 orders of magnitude in pressure with sub-second response}

\author{L.J. Bannenberg}
\affiliation{Faculty of Applied Sciences, Delft University of Technology, Mekelweg 15, 2629 JB Delft, The Netherlands}
\email{l.j.bannenberg@tudelft.nl}
\author{H. Schreuders}
\affiliation{Faculty of Applied Sciences, Delft University of Technology, Mekelweg 15, 2629 JB Delft, The Netherlands}
\author{B. Dam}
\affiliation{Faculty of Applied Sciences, Delft University of Technology, Mekelweg 15, 2629 JB Delft, The Netherlands}


\begin{abstract}
Hydrogen detection in a reliable, fast, and cost-effective manner is a prerequisite for the large-scale implementation of hydrogen in a green economy. We present thin film Ta$_{1-y}$Pd$_y$ as effective optical sensing materials with extremely wide sensing ranges covering at least seven orders of magnitude in hydrogen pressure. Nanoconfinement of the Ta$_{1-y}$Pd$_y$ layer suppresses the first-order phase transitions present in bulk and ensures a sensing response free of any hysteresis. Unlike other sensing materials, Ta$_{1-y}$Pd$_y$ features the special property that the sensing range can be easily tuned by varying the Pd concentration without a reduction of the sensitivity of the sensing material. Combined with a suitable capping layer, sub-second response times can be achieved even at room temperature, faster than any other known thin-film hydrogen sensor. 

\end{abstract}
\maketitle
\section{Introduction}
Hydrogen is projected to fuel tomorrow's green economy \cite{brandon2017,natmat2018,abe2019,glenk2019,bannenberg2020review,el2020}. 
Reliable and inexpensive hydrogen sensors are a vital ingredient for the large scale adaptation of hydrogen, being an odorless, colorless and tasteless gas with a wide flammability range in air. As such, any leaks of hydrogen must be detected immediately, calling for efficient, fast, reliable and cost-effective hydrogen sensors. Apart from safety applications, precise monitoring of hydrogen concentrations and pressures is crucial for the efficient and reliable operation of hydrogen fuel cells, CO$_2$ conversion devices, and in a variety of industrial processes \cite{bakenne2016,el2020}.

Optical metal-hydride hydrogen sensors are an attractive candidate for large scale implementation in the future hydrogen economy \cite{bannenberg2020,darmadi2020,sterl2020,koo2020,chen2021}. Their working principle is based on the fact that the optical properties change when metal hydrides partly hydrogenate when they are exposed to a hydrogen atmosphere. These changes are probed by for example measuring the fraction of transmitted or reflected light or the frequency shift of the (localized) surface plasmon resonance peak and from one of these optical signals the partial hydrogen pressure $P_{H2}$ can be determined. Compared to conventional ways of detecting hydrogen such as catalytic resistor detectors and electrochemical devices, optical fiber hydrogen sensors are inherently safe and can be made small and inexpensive \cite{hughes1991,butler1991,hubert2011,perrotton2011,slaman2012,javahiraly2014,bannenberg2020,darmadi2020,koo2020,chen2021}. 

An ideal hydrogen sensing material should feature a large sensing range, hysteresis-free response, high sensitivity and fast response to changing hydrogen pressures. Palladium, being the metal hydride considered most frequently for hydrogen sensing applications \cite{hughes1991,butler1991,silva2012,perrotton2013,wadell2014}, features some of these properties: it is capable of dissociating hydrogen at room temperature and has a modest sensing range of three orders of magnitude in pressure. However, the first-order transition from the dilute $\alpha$-PdH$_x$ to the higher hydrogen concentration PdH$_x$ $\beta$-phase occurring upon hydrogenation causes relatively long response times and renders the optical response highly hysteric and non-linear. Alloying Pd with elements including Au \cite{zhao2006,westerwaal2013,wadell2015,nugroho2018,nugroho2019,bannenberg2019PdAu,palm2019}, Ni\cite{dai2012,dai2014},  Ag \cite{wang2007} and Cu \cite{darmadi2019} partly overcomes these serious shortcomings: it eliminates the first order phase transition but also significantly reduces the optical contrast and thus reduces the sensitivity of the sensor. Moreover, hysteresis arising from the expansion of the film upon hydrogenation and its clamping of the film to the support remains present \cite{bannenberg2019PdAu}. Alternatively, hydrogen sensors with separate hydrogen dissociation and sensing functionality can be developed \cite{slaman2007,boelsma2017,bannenberg2019}. For example, Pd-capped Hf thin films were shown to be able to hysteresis-free sense hydrogen over a large pressure range. However, its functionality has only been demonstrated at elevated temperatures ($T$ $\geq$ 90$\degree$C) and modest hydrogen pressures of $P_{H2}$ $<$ 10$^2$~Pa. 

As such, new sensing materials have to be discovered to simultaneously meet the requirements of a large sensing range, hysteresis-free response, high sensitivity and fast response to changing hydrogen pressures both at room and elevated temperatures. In terms of material properties, this can be realized by a large solubility of hydrogen within one thermodynamic phase, a profound change of the optical properties upon hydrogenation and a high diffusivity of hydrogen \cite{fukai2006,bannenberg2020}. Tantalum is in principle a suitable candidate: TaH$_x$ has a large solubility window of 0 $<$ $x$ $\lesssim$ 0.9 for $T$ $>$ 61~$\degree$C within one thermodynamic phase that would thus result in a wide sensing range without hysteresis. Furthermore, the optical properties of Ta show a relatively large and gradual change upon hydrogenation, thus ensuring a high sensitivity of the sensor \cite{bannenberg2019} while the body-centered cubic structure of Ta results in a high hydrogen diffusivity that allows fast response times \cite{bannenberg2019}. However, below its critical temperature of $T_C$ = 61~$\degree$C a series of phase transitions  reported for bulk Ta and associated nucleation of domains render the hydrogenation slow and, most importantly, highly hysteric, which makes Ta in principle unsuitable for sensing purposes.


Using a combination of nanoconfinement and alloying, we tailor the properties of Ta and design hysteresis-free hydrogen sensors with extremely wide sensing ranges of at least seven orders of magnitude in hydrogen pressure. The nanoconfinement of Ta$_{1-y}$Pd$_y$ suppresses the first-order phase transitions present in bulk at room temperature and ensures a sensing response free of any hysteresis. Unlike other sensing materials, the sensing range of Ta$_{1-y}$Pd$_y$ can be easily tuned by varying the Pd concentration in the alloy without a reduction of the optical contrast and thus the sensitivity of the sensing material. Combined with a capping layer of Pd$_{0.6}$Au$_{0.35}$Cu$_{0.05}$ covered with a 30~nm polytetrafluoroethylene (PTFE) layer, it features a sub-second response at room temperature in a pressure range of 10$^{+2}$ $<$ $P_{H2}$ $<$ 10$^{+5}$, i.e. a concentration of 0.1 to 100\% under ambient conditions. As such, the Ta$_{1-y}$Pd$_y$ based thin film sensor meets the most stringent set of criteria for hydrogen sensors of the U.S. Department of Energy \cite{doe}, which, combined with its relatively easy and cheap manufacturing, make it exceptionally well-suited for the large scale adaptation in tomorrow's hydrogen economy.

\section{Results and discussion}

\subsection{Optical response}
A suitable optical hydrogen sensing material should feature a monotonic and gradual change of the optical properties in a large range in hydrogen pressures without any dependence on the (pressure) history of the sensor. To study the behavior of the Ta$_{1-y}$Pd$_y$ thin films, we measure the changes of the white-light optical transmission $\mathcal{T}$ relative to that of the as-prepared state $\mathcal{T}_{prep}$ when applying a series of increasing and decreasing hydrogen pressure steps at room temperature.

\begin{figure}[h!]
\begin{center}
\includegraphics[width= 0.45\textwidth]{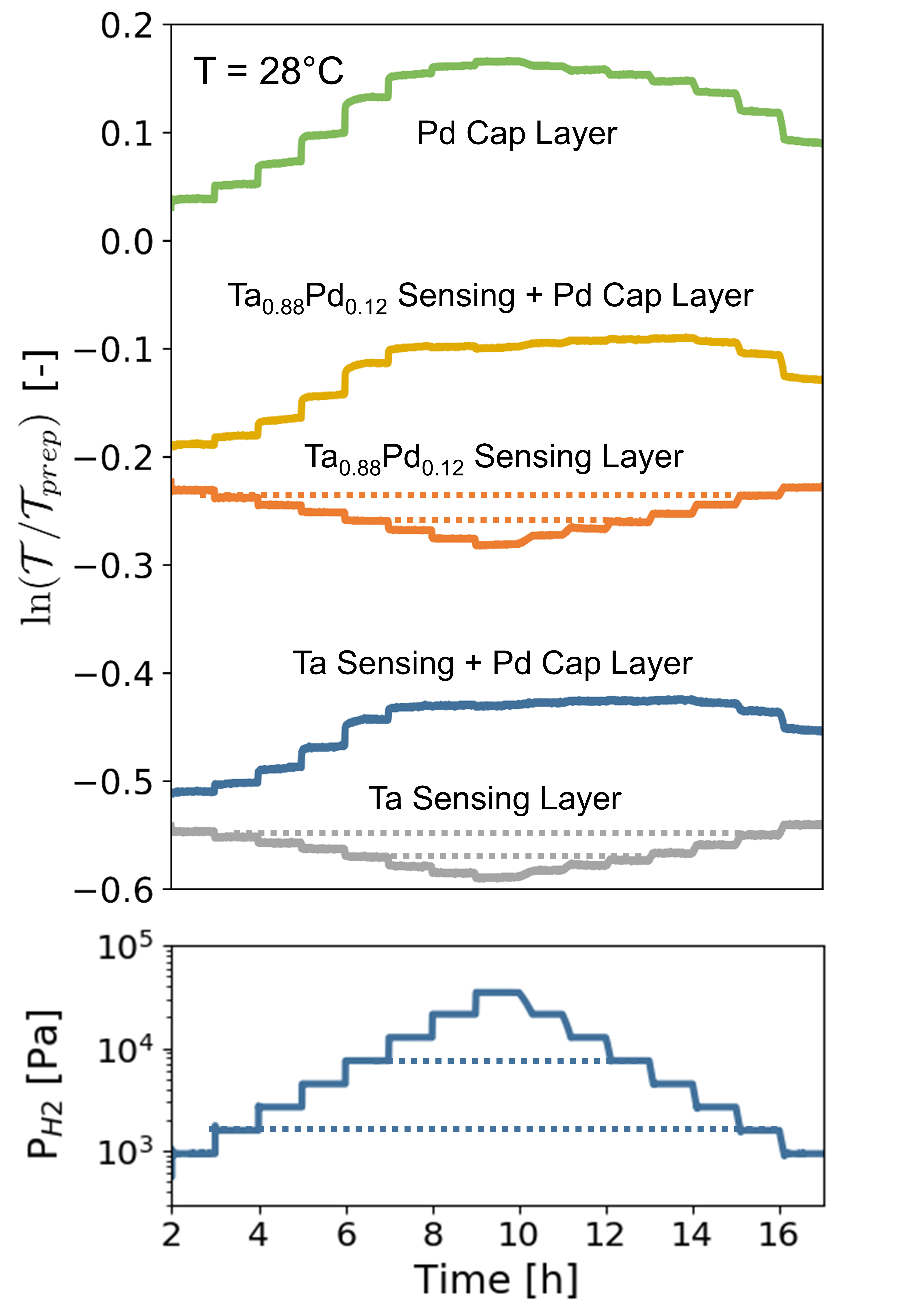}
\caption{Changes of the white light optical transmission $\mathcal{T}$ of a Ta and Ta$_{0.88}$Pd$_{0.12}$ thin film both capped with 10~nm Pd and with a 4~nm Ti adhesion layer a function of time as measured relative to the transmission of the as-prepared film ($\mathcal{T}_{prep}$). It includes both the response of the uncorrected samples, i.e. the contribution of the sensing and cap layer, and the corrected samples, i.e. the contribution of the sensing layer computed by subtracting the contribution of the Pd cap layer from the uncorrected sample. The contribution from the Pd cap layer is determined by measuring a Pd-capped Ta$_{0.5}$Pd$_{0.5}$ thin film. Ta$_{0.5}$Pd$_{0.5}$ does not show any optical response in the investigated pressure range and has similar adhesion conditions as the Ta$_{1-y}$Pd$_y$ of interest, ensuring that the optical response of the Pd layer mimics the one on top of the film of interest. The films were exposed to various increasing and decreasing pressure steps between $P_{H2}$ = 1$\times$10$^{+3}$ - 4$\times$10$^{+4}$ Pa at $T$ = 28~$\degree$C. The dashed lines indicate levels of the same transmission (top panel) and pressure (bottom panel).}
\label{stepresponse}
\end{center}
\end{figure}

\begin{figure*}[tb]
\begin{center}
\includegraphics[width= 1\textwidth]{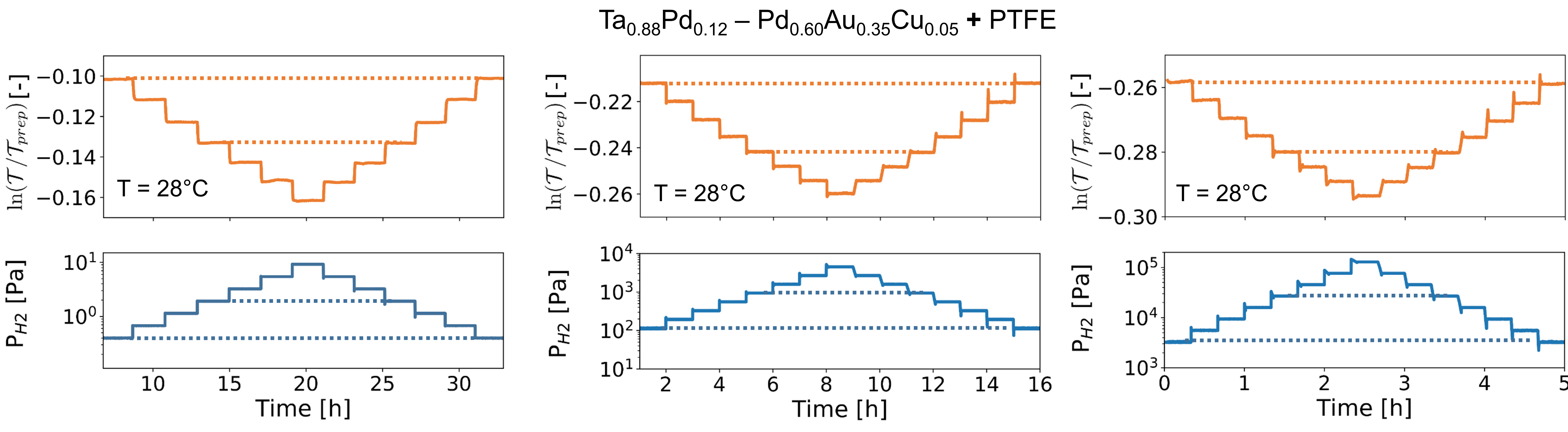}
\caption{Changes of the white light optical transmission $\mathcal{T}$ of a 40~nm Ta$_{0.88}$Pd$_{0.12}$ thin film with a 4~nm Ti adhesion layer capped with a 10~nm Pd$_{0.60}$Au$_{0.35}$Cu$_{0.05}$ layer that is covered with a 30~nm PTFE layer (`Ta$_{0.88}$Pd$_{0.12}$  - Pd$_{0.60}$Au$_{0.35}$Cu$_{0.05}$ + PTFE'). The optical transmission is measured as a function of time and relative to the transmission of the as-prepared film ($\mathcal{T}_{prep}$). The film was exposed at $T$ = 28~$\degree$C to various increasing and decreasing pressure steps of (a) 4.0$\times$10$^{-1}$ $\leq$ $P_{H2}$ $\leq$ 9.2$\times$10$^{+0}$~Pa, (b) 1.2$\times$10$^{+2}$ $\leq$ $P_{H2}$ $\leq$ 4.5$\times$10$^{+3}$~Pa and (c) 3.5$\times$10$^{+3}$ $\leq$ $P_{H2}$ $\leq$ 1.5$\times$10$^{+5}$~Pa. The dashed lines indicate levels of the same transmission (top panel) and pressure (bottom panel).}
\label{TaPd_response_PTFE}
\end{center}
\end{figure*}

\begin{figure*}[tb]
\begin{center}
\includegraphics[width= 1\textwidth]{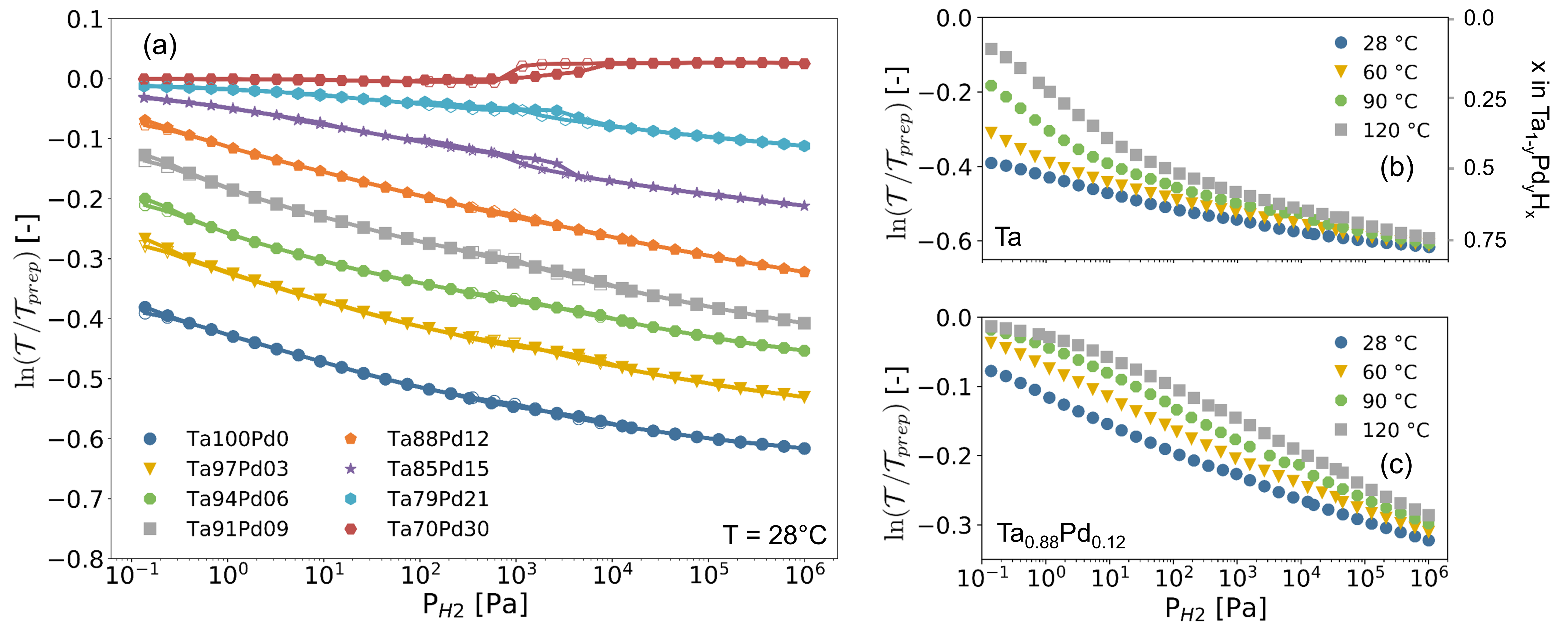}
\caption{Partial hydrogen pressure and temperature dependence of the optical transmission $\mathcal{T}$ of 40~nm Ta$_{1-y}$Pd$_y$ sensing layers measured relative to the optical transmission of the as-prepared state ($\mathcal{T}_{prep}$). The measurements have been performed using the same method as in Figure \ref{stepresponse}. Each data-point corresponds to the optical transmission after exposing the film for at least one hour to a constant pressure of $P_{H2}$ = 10$^{-1}$ - 10$^{+6}$~Pa, where the closed data points correspond to increasing pressure steps, and the open ones to decreasing pressure steps. Panel (a) presents the pressure dependence for different values of $y$ in Ta$_{1-y}$Pd$_y$ at $T$ = 28~$\degree$C. Panel (b) and (c) present the pressure dependence of the changes in optical transmission for (b) Ta and (c) Ta$_{0.88}$Pd$_{0.12}$ at different temperatures. The right axis indicates $x$ in Ta$_{1-y}$Pd$_y$H$_x$ as based on the scaling between the optical response and the hydrogen content of Ta obtained using \textit{in-situ} neutron reflectometry measurements \cite{bannenberg2019}.}
\label{TaPd_response}
\end{center}
\end{figure*}

\begin{figure}[tb]
\begin{center}
\includegraphics[width= 0.45\textwidth]{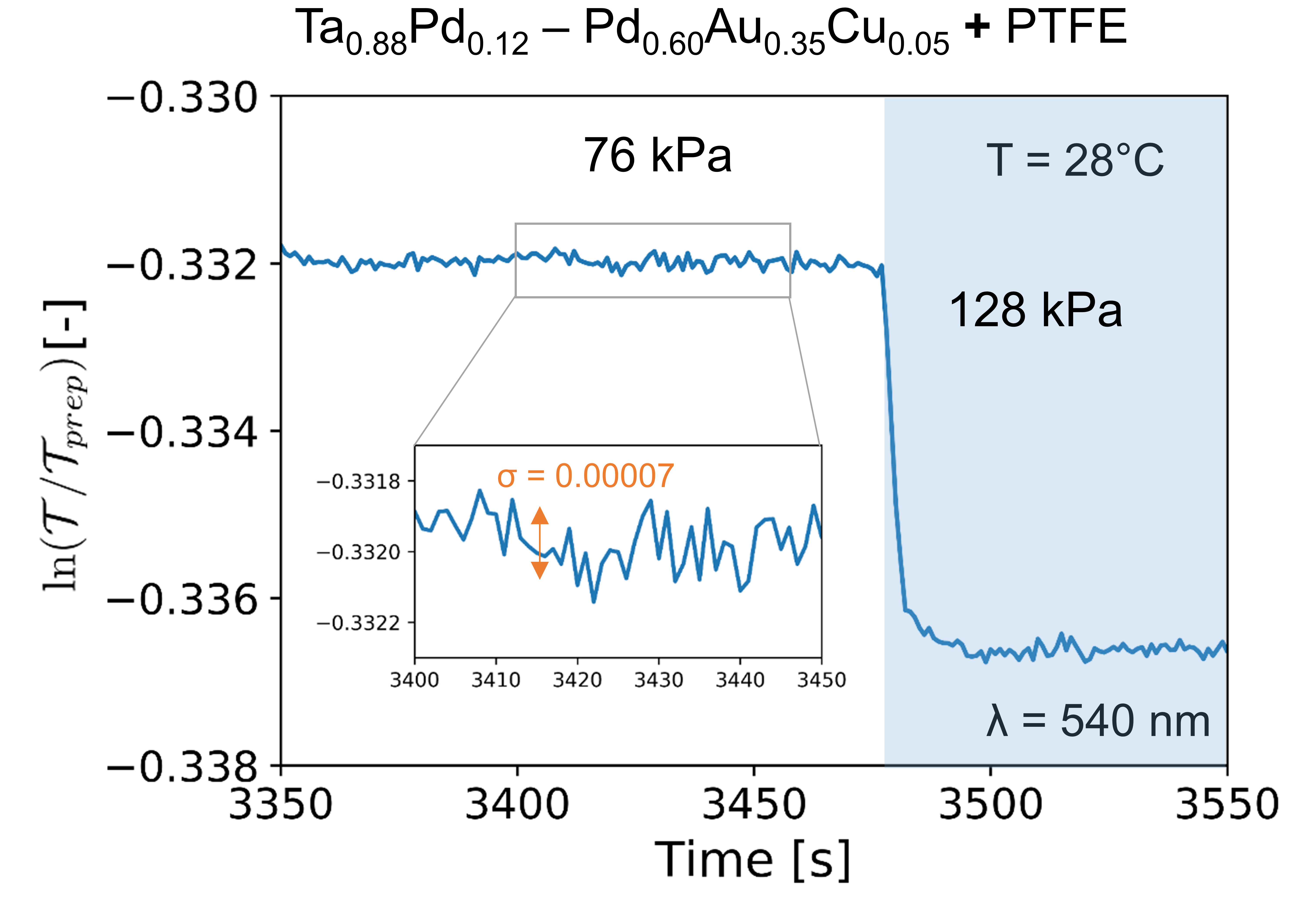}
\caption{Changes of the optical transmission $\mathcal{T}$ of 40~nm Ta$_{0.88}$Pd$_{0.12}$ thin film with a 4~nm Ti adhesion layer and capped with a 10~nm Pd$_{0.60}$Au$_{0.35}$Cu$_{0.05}$ layer that is covered with a 30~nm PTFE layer as a function of time (`Ta$_{0.88}$Pd$_{0.12}$  - Pd$_{0.60}$Au$_{0.35}$Cu$_{0.05}$ + PTFE'). At $t$ = 1780~s, the pressure is changed from $P_{H2}$ = 3,300 to 5,500~Pa at a temperature of $T$ = 28~$\degree$C. The standard deviation of the optical transmission is $\sigma$ = 0.00007 at a sampling frequency of 1~Hz and a wavelength $\lambda$ = 540~nm. This results in a sensitivity of $\Delta P_{H2}$/$P_{H2}$ = 0.008.}
\label{Resolution}
\end{center}
\end{figure}

\begin{figure}[tb]
\begin{center}
\includegraphics[width= 0.38\textwidth]{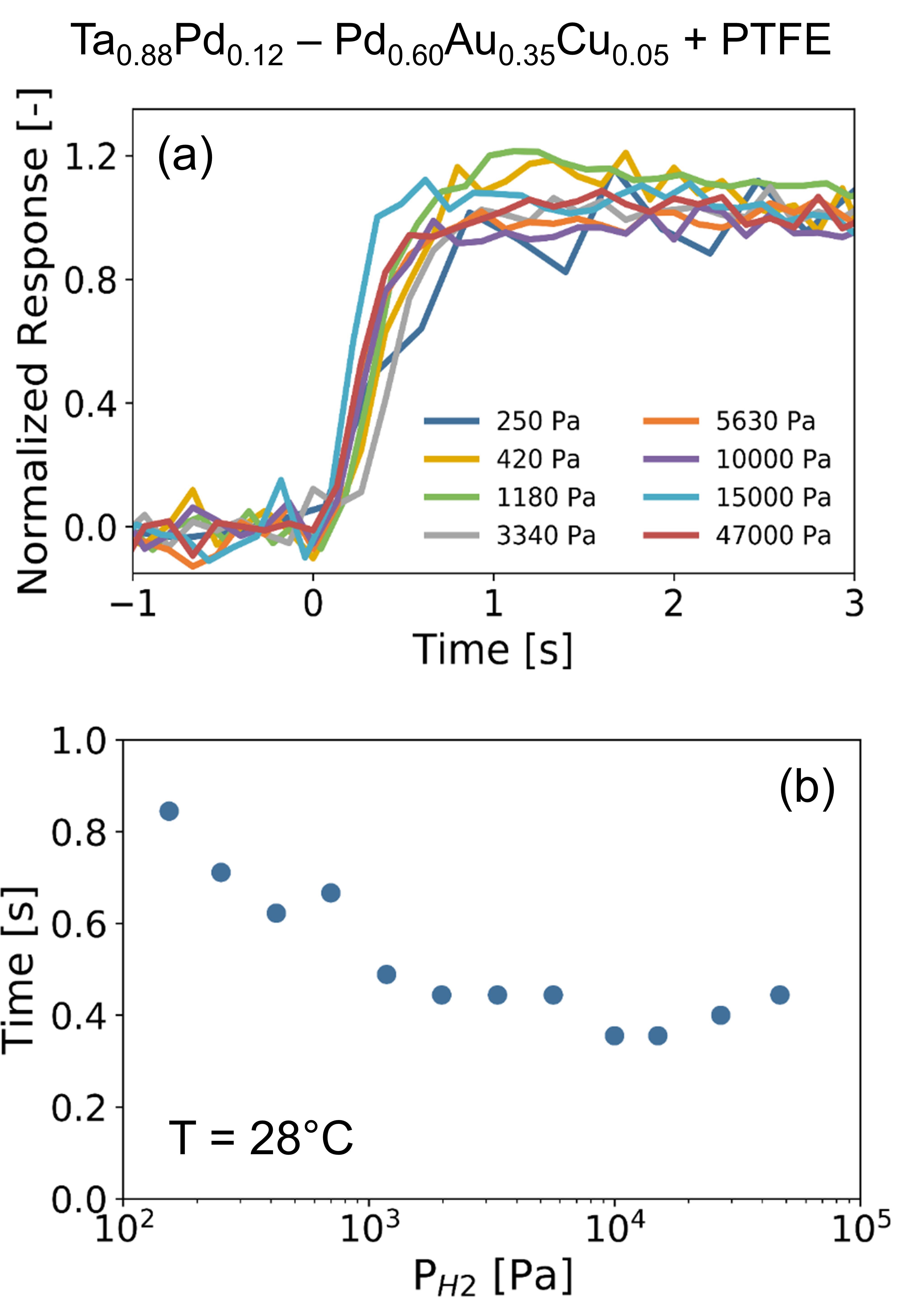}
\caption{Absorption kinetics of a 40~nm Ta$_{0.88}$Pd$_{0.12}$ thin film with a 4~nm Ti adhesion layer capped with a 10~nm Pd$_{0.60}$Au$_{0.35}$Cu$_{0.05}$ layer that is covered with a 30~nm PTFE layer (`Ta$_{0.88}$Pd$_{0.12}$  - Pd$_{0.60}$Au$_{0.35}$Cu$_{0.05}$ + PTFE') at $T$ = 28~$\degree$C. (a) Normalized responses of the thin film to a series of pressure steps between $P_{H2}$ = 0.5 10$^2$~Pa and the partial hydrogen pressure indicated. (b) Hydrogen pressure dependence of the response time of the thin film. The response time is defined as the time to reach 90\% of the total signal.}
\label{ResponseTime}
\end{center}
\end{figure}

\begin{figure}[tb]
\begin{center}
\includegraphics[width= 0.40\textwidth]{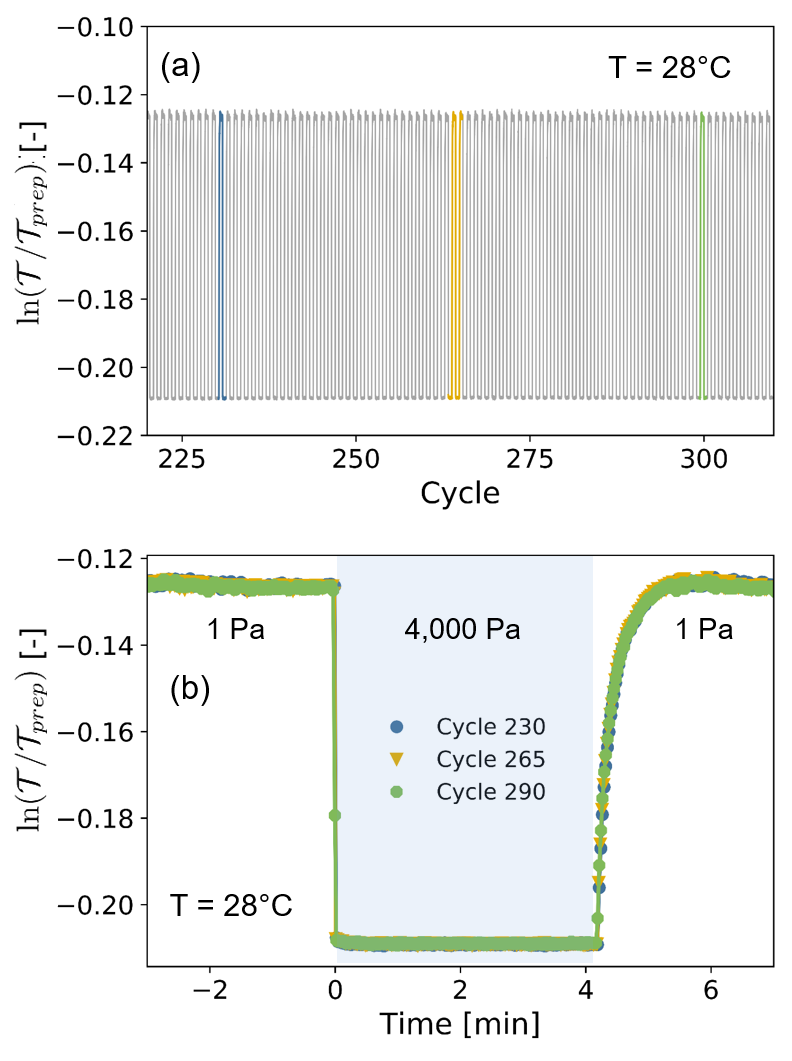}
\caption{Stability of a 40~nm Ta$_{0.88}$Pd$_{0.12}$ thin film with a 4~nm Ti adhesion layer capped with a 10~nm Pd$_{0.60}$Au$_{0.35}$Cu$_{0.05}$ layer that is covered with a 30~nm PTFE layer (`Ta$_{0.88}$Pd$_{0.12}$  - Pd$_{0.60}$Au$_{0.35}$Cu$_{0.05}$ + PTFE'). The transmission is measured relative to the optical transmission of the as-prepared state ($\mathcal{T}_{prep}$). (a) Optical (white light) response of the film to 110 cycles in which the pressure was varied between partial hydrogenation ($P_{H2}$ = 4.0 10$^{+3}$~Pa) and partial dehydrogenation ($P_{H2}$ = 1~Pa) at $T$ = 28~$\degree$C. (b) The identical behavior of three individual hydrogenation cycles selected at random from (a). Note that the relatively slow desorption mainly originates from the sluggish removal of hydrogen from the measurement cell.}
\label{Stability}
\end{center}
\end{figure}

\begin{figure*}[tb]
\begin{center}
\includegraphics[width= 0.75\textwidth]{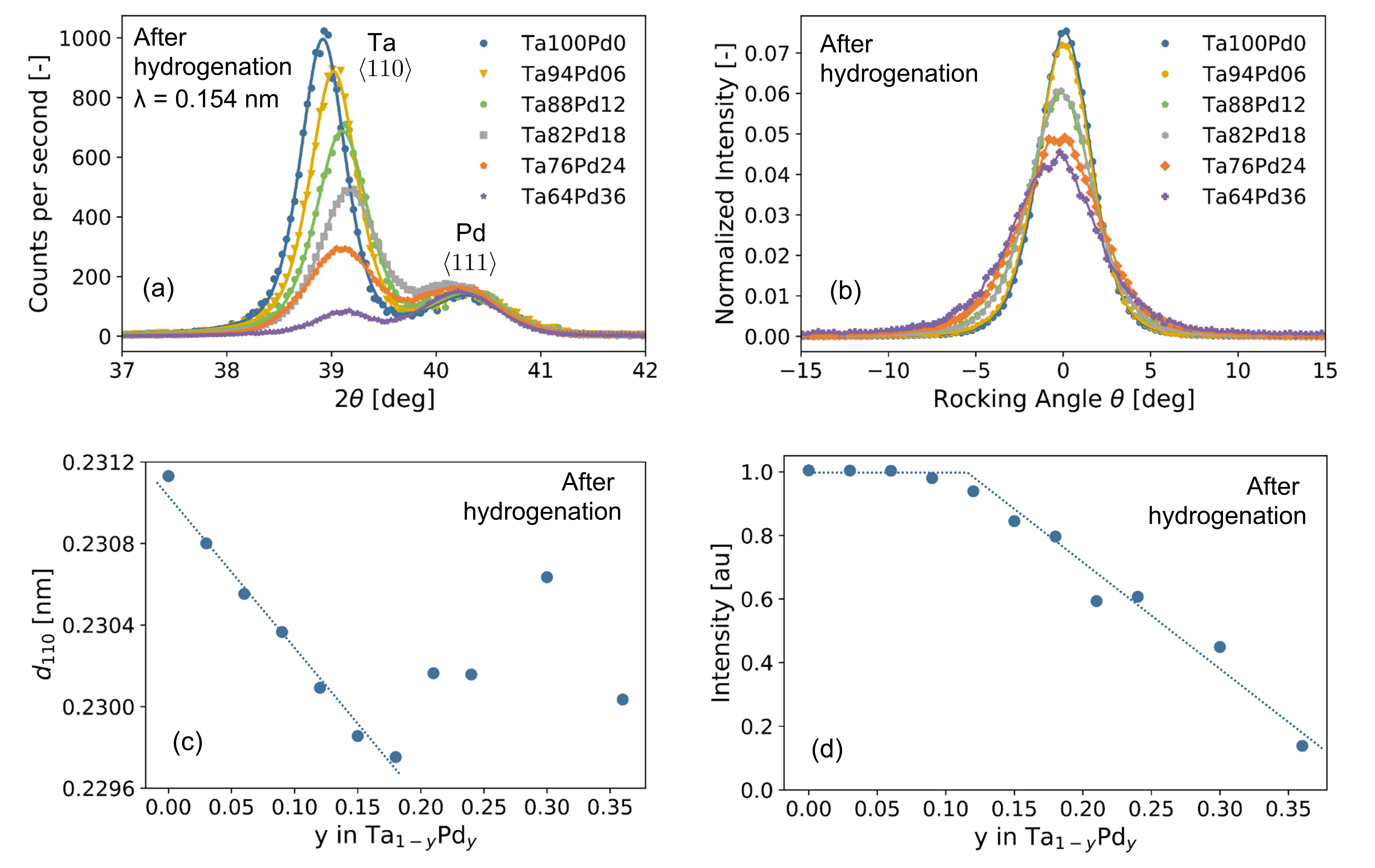}
\caption{Ex-situ X-ray Diffraction (XRD) results of the 40~nm Ta$_{1-y}$Pd$_{y}$ thin films with a 4~nm Ti adhesion layer and capped with a 10~nm Pd layer after exposure of the thin films to hydrogen and measured in air. (a) Diffractograms (Cu-K$\alpha$, $\lambda$ = 0.1542~nm) of the Ta$_{1-y}$Pd$_{y}$ thin films. The continuous lines represent fits of two pseudo-Voigt functions to the experimental data. (b) Rocking curves of the Ta$_{1-y}$Pd$_{y}$ thin films around the Ta$_{1-y}$Pd$_{y}$ $\langle110\rangle$ peak. (c) Pd doping dependence of the $d_{110}$-spacing in Ta$_{1-y}$Pd$_{y}$.  (d) Pd concentration dependence of the total intensity of the $\langle$110$\rangle$ diffraction peak in Ta$_{1-y}$Pd$_{y}$ in which the effect of both the changing amplitude and width are incorporated. The intensity is scaled to the intensity of the Ta sample. The dashed lines serve as guides to the eye.}
\label{ExSituXRD}
\end{center}
\end{figure*}

\begin{figure*}[tb]
\begin{center}
\includegraphics[width= 0.75\textwidth]{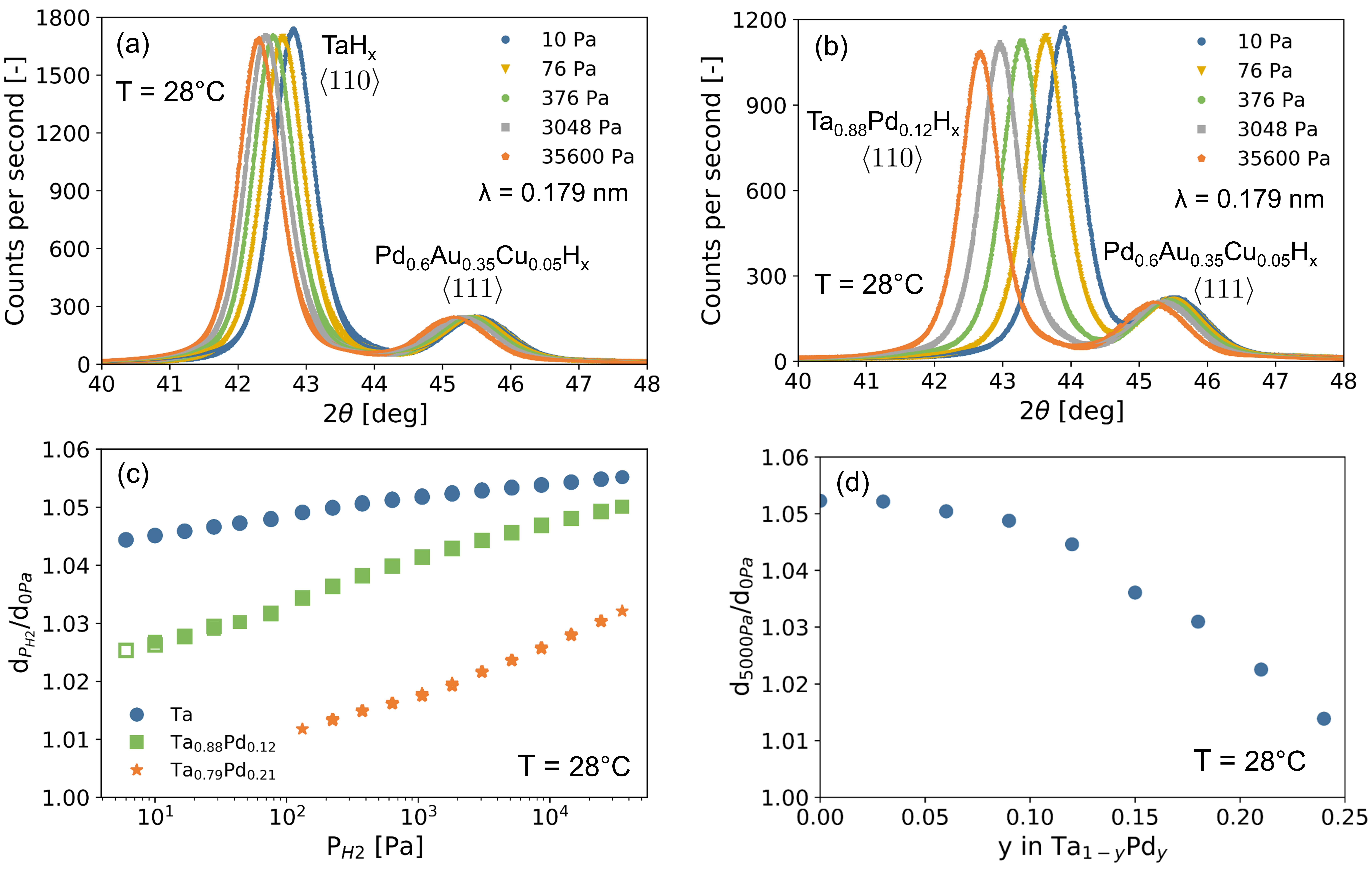}
\caption{In-situ XRD results of the 40~nm Ta$_{1-y}$Pd$_{y}$ thin films with a 4~nm Ti adhesion layer and capped with a 10~nm Pd$_{0.6}$Au$_{0.35}$Cu$_{0.05}$ layer that is covered with a 30~nm PTFE layer for $T$ = 28~$\degree$C. (a,b) Diffractograms (Co-K$\alpha$ $\lambda$ = 0.179~nm) of the Ta$_{1-y}$Pd$_{y}$ thin films with (a) $y$ = 0 and $y$ = 0.12 measured for the hydrogen pressures indicated in the legend and for increasing pressure steps. The continuous lines represent fits of two pseudo-Voigt functions to the experimental data. (c) Hydrogen pressure dependence of the expansion of the $d_{110}$-spacing in Ta$_{1-y}$Pd$_{y}$ relative to the unloaded state as measured by both increasing (closed symbols) and decreasing the pressure (open symbols). (d) Doping dependence of the expansion of the $d_{110}$-spacing measured at $P_{H2}$ = 5.0 10$^{+3}$~Pa.}
\label{InSituXRD}
\end{center}
\end{figure*}

The results for the Pd capped 40~nm Ta and Ta$_{0.88}$Pd$_{0.12}$ thin films are displayed in {Figure \ref{stepresponse}} and reveal a complicated optical response: the response of these films is highly non-linear and hysteretic as a result of interference between the optical response of the Pd-capping layer and the Ta$_{1-y}$Pd$_y$ sensing layer. To deconvolute the response of the Pd capping layer and the Ta$_{1-y}$Pd$_y$ sensing layers, we subtract the optical response of a Pd-capped Ta$_{0.5}$Pd$_{0.5}$ thin film from the film of interest (Fig. \ref{stepresponse}). Ta$_{0.5}$Pd$_{0.5}$ does not show any optical response in the investigated pressure range to hydrogen and provides similar adhesion conditions to the Pd cap layer, ensuring that the response of the Pd layer to hydrogen mimics the one on top of the film of interest. The resulting deconvolution displays a remarkable result. The non-linear and hysteretic optical response to hydrogen of the Pd capped Ta$_{1-y}$Pd$_y$ thin films is completely the result of the Pd capping layer. Indeed, the optical transmission of the Ta$_{1-y}$Pd$_y$ sensing layer decreases monotonically with increasing pressure, is stable, well-defined and free of any hysteresis. 

To obtain a hysteresis-free monotonic response without the need of such a deconvolution we need a different capping layer for which we select Pd$_{0.60}$Au$_{0.35}$Cu$_{0.05}$. The introduction of Au and Cu in this alloy ensures that this layer hardly hydrogenates, thus not generating a sizable optical contrast (Figure \ref{OpticalResponseCap}). Moreover, the source of the hysteresis, the first-order metal-to-metal hydride transition in Pd, is completely suppressed \cite{sakamoto1991,darmadi2019}. In addition, we cover this capping layer by a 30~nm PTFE layer to promote the hydrogenation kinetics \cite{ngene2014,nugroho2019}. {Figure \ref{TaPd_response_PTFE}} displays the changes in white-light optical transmission of the Ta$_{0.88}$Pd$_{0.12}$ thin film capped with 10 nm of Pd$_{0.6}$Au$_{0.35}$Cu$_{0.05}$ in three different pressure regions with 4.0$\times$10$^{-1}$ $<$ $P_{H2}$ $<$ 1.5$\times$10$^{+5}$~Pa. In all three pressure regions, the optical transmission shows a substantial and monotonic decrease with increasing pressure. The levels of transmission are well-defined and stable for a given partial hydrogen pressure, and, importantly, free of any hysteresis: the optical transmission is, in accordance with the \textit{in-situ} XRD measurements (See Section 2.2), the same after increasing and decreasing pressure steps. We note that the sharp peaks in optical transmission during desorption in Fig. \ref{TaPd_response_PTFE} are the results of undershoots in hydrogen pressure during the measurements. As such, these peaks are exemplary of the fast response of the material to changing hydrogen pressures.

In {Figure \ref{TaPd_response}} we summarize the optical transmission measurements performed for a set of different pressures between 1.0$\times$10$^{-1}$ $<$ $P_{H2}$ $<$ 1.0$\times$10$^{+6}$~Pa, for various Pd/Ta ratios of 0.0 $\leq$ $y$ $\leq$ 0.36, and for temperatures of 28, 60, 90 and 120~$\degree$C. In this figure, we plot the pressure-transmission-isotherms (PTIs) of the sensing layers, where each closed data-point corresponds to the optical transmission obtained after exposing the film for at least one hour to a constant pressure after an increase in pressure, and the open points to the transmission after decreasing pressure steps.

The PTIs at room temperature (Figure \ref{TaPd_response}(a)) reveal for all Ta$_{1-y}$Pd$_y$ films with $y$ $\lesssim$ 0.12 including pure Ta an excellent optical response with no hysteresis and a large and almost constant sensitivity over an extremely wide sensing range of at least 7 orders of magnitude in pressure. This makes this material unique, as the metal hydride known so far with the largest sensing range at room-temperature, Pd$_{1-y}$Au$_y$, only features a sensing range of roughly 4 orders in magnitude. 

The hydrogenation of metal-hydrides depends in general on the temperature, with the result that the PTIs and thus the sensing range shifts to higher pressures with increasing temperatures. Nevertheless, we find that for all temperatures measured the Ta$_{1-y}$Pd$_y$ thin films with $y$ $\lesssim$ 0.12 keep their sensing-range of at least 7 orders in magnitude between 10$^{-1}$ $<$ $P_{H2}$ $<$ 10$^{+6}$~Pa, ensuring that the sensing material can be used for a wide range of applications and under various operating conditions. Differently, the films with $y$ $\gtrsim$ 0.12 display a more complex optical response to hydrogen. As this response involves substantial hysteresis in the optical response, these alloys are unsuitable for most hydrogen sensing applications. 



In contrast to many other sensing materials such as Pd$_{1-y}$Au$_y$, Pd$_{1-y}$Ag$_y$ and Mg-alloys, Ta$_{1-y}$Pd$_y$ has the major benefit that the compositional tuning of the sensing range does not decrease the optical contrast and thus the sensitivity of the sensor. Despite the fact that the hysteresis-free sensing range in Ta$_{1-y}$Pd$_y$ can be tuned by varying $y$ between 0 $<$ $y$ $<$ 0.12, we envision that Ta$_{0.88}$Pd$_{0.12}$ is the most attractive composition for the majority of applications. It has an exceptionally large sensing range of at least 1.0 $10^{-1}$ $<$ $P_{H2}$ $<$ 1.0 $10^{+6}$~Pa, i.e. seven orders in magnitude, for a wide temperature window ranging from room temperature up to $T$ = 120~$\degree$C (Figure \ref{TaPd_response}(c)). This massive sensing range is larger than any other material reported so far and highlights the versatile nature of the material: it can be used both to discover tiny hydrogen leaks at extremely low hydrogen pressures as well as to probe hydrogen concentrations in environments where hydrogen is omnipresent such as in fuel cells or in various industrial processes.

Furthermore, Ta$_{0.88}$Pd$_{0.12}$ has the largest sensitivity among all compounds at room temperature in the range 1.0 $10^{+3}$ $<$ $P_{H2}$ $<$ 1.0 $10^{+5}$~Pa, i.e., the range most applicable for hydrogen leak detection. In fact, wavelength-dependent measurements reveal that the optical contrast, and thus the sensitivity, is strongly wavelength dependent (Figure \ref{WavelengthDependence}). Its sensitivity is largest for wavelengths of 470 $\lesssim$ $\lambda$ $\lesssim$ 550~nm. Figure \ref{Resolution} illustrates that in combination with a sampling frequency of 1~Hz and at a wavelength of $\lambda$ = 540~nm we may achieve a resolution of $\Delta P_{H2}$/$P_{H2}$ = 0.008 that is relatively constant with pressure, i.e. substantially better than the most stringent requirement of 5\% set by the U.S. Department of Energy \cite{doe}.

The response time of a hydrogen sensor is a key indicator of its performance, especially for safety applications and in particular for hydrogen pressures relatively close to the explosive limit of 4\% in air. Figure \ref{ResponseTime} presents the room temperature response time measurements for the 40~nm Ta$_{0.88}$Pd$_{0.12}$ thin film capped with a 10~nm Pd$_{0.6}$Au$_{0.35}$Cu$_{0.05}$ layer that is covered with a 30~nm PTFE layer. In Fig. \ref{ResponseTime}(a) we display the normalized responses of the thin film to a series of increasing pressure steps and in Fig. \ref{ResponseTime}(b) we depict the pressure dependence of the response time, defined as the time to reach 90\% of the total signal in Fig. \ref{ResponseTime}(a).  While other thin films reported in the literature often feature relatively long response times, especially at room temperature and for Pd-based sensors, the room temperature response time of the Ta$_{0.88}$Pd$_{0.12}$ based thin film is below 1~s in the entire pressure window of $P_{H2}$ = 10$^{+2}$ $<$ $P_{H2}$ $<$ 10$^{+5}$, i.e. a hydrogen concentration range of 0.1 to 100\% under ambient conditions. Apparently, we benefit from the the well-know high diffusivity of hydrogen in bcc Ta \cite{fukai2006}. In fact, the response time is merely limited by the amount of hydrogen that is dissociated at the surface of the capping layer and not by the hydrogen diffusion through the sensing or capping layer \cite{bannenberg2020response}. As such, even faster responses can be achieved by reducing the thickness of the Ta$_{0.88}$Pd$_{0.12}$ sensing layer as it reduces the total amount of hydrogen that needs to be dissociated at the surface. Note, that the response times we achieve with our 40~nm sensing layer are already shorter than other hydrogen sensors including metal-hydride nanoparticles \cite{nugroho2019}.

Another key requirement of hydrogen sensors is long-term stability of the optical response. To illustrate the stability of the sensor response, we exposed the thin film to 310 cycles of hydrogen between $P_{H2}$ = 1.0 and 4.0 10$^{+3}$~Pa at $T$ = 28~$\degree$C (only 90 cycles displayed). Figure \ref{Stability}(a) shows that the cycles are identical to each other, even after exposure to over 300 cycles of hydrogen. This conclusion is further underscored by the three individual cycles selected at random for which we observe a very close correspondence (Figure \ref{Stability}(b)). Together, these results indicate an excellent stability and reproducibility of the optical response of the thin film sensor for a large number of cycles. 

To ensure a long lifetime and stable performance of the hydrogen sensor, it should have a good chemical selectivity of the material and protection against poisoning by chemical species. Although the inclusion of 5\% of Cu in the capping layer of Pd provides protection against deactivation by CO \cite{darmadi2019}, additional protection can be provided by coating the PTFE layer with 30~nm of poly(methylmethacrylate) (PMMA). This coating has been shown to provide significant protection against CO, CO$_2$ and NO$_2$ without a reduction of the response times \cite{nugroho2019}.

\subsection{Structural behavior}
The sensing results on Ta$_{1-y}$Pd$_y$ thin films suggests that (i) no phase segregation occurs and thus that a solid solution of Ta and Pd is formed and that no segregation takes place during hydrogenation as this would reduce the lifetime and stability of the sensor and (ii) that no (first-order) phase transitions occur upon hydrogenation as this would induce a hysteretic response with long response times. 

X-Ray Diffraction (XRD) measurements reveal that the Pd-capped Ta$_{1-y}$Pd$_y$ thin films form, as in bulk \cite{darby1963,waterstrat1978}, a solid solution with no signs of phase segregation for $y$ $\lesssim$ 0.12. This holds true both for the as-prepared films and the films after exposure to at least 300 cycles of hydrogen. The measurements are displayed in Figure \ref{ExSituXRD} and show that all Ta$_{1-y}$Pd$_y$ thin films are textured in the $\langle110\rangle$ direction (Figure \ref{ExSituXRD}(a,b) and \ref{Rocking_before_after_H2}). With increasing Pd doping, the $\langle$110$\rangle$ diffraction peak slightly broadens and shifts to the right for $y$ $\lesssim$ 0.21, corresponding to an almost linear decrease of the $d_{110}$-spacing with increasing Pd concentration (Figure \ref{ExSituXRD}(c)). This contraction of the unit cell is anticipated owing to the lower atomic number of Pd as compared to Ta and has the benefit that it would shift the pressure range of the sensing material to higher pressures. Looking at the total intensity of the diffraction peaks, the measurements show an almost constant intensity for $y$ $\lesssim$ 0.12, consistent with the fact that the films are a solid solution of Ta and Pd. For $y$ $\gtrsim$ 0.12 a continuous decrease of the intensity of the $\langle110\rangle$ diffraction peak is observed with increasing Pd concentration (Figure \ref{ExSituXRD}(d)), and ultimately, for Ta$_{0.5}$Pd$_{0.5}$ no diffraction peaks related to a Ta$_{1-y}$Pd$_y$ phase are detected. As other crystalline phases may be textured as well and have a different preferential orientation, we also performed diffraction measurements at different tilting angles of the film (Figure \ref{ChiXRD}). Unlike bulk Ta$_{1-y}$Pd$_y$ \cite{darby1963,waterstrat1978}, these measurements did not reveal any additional crystalline phase. As such, the XRD measurements are consistent with the formation of a solid solution of Ta and Pd for $y$ $\lesssim$ 0.12 and the coexistence of a crystalline phase and an (X-ray) amorphous phase for $y$ $\gtrsim$ 0.12 of which the amorphous fraction increases with increasing Pd. The formation of a solid solution which is stable on hydrogen cycling explains why the films with $y$ $\lesssim$ 0.12 are most suitable for hydrogen sensing purposes.

While the hydrogenation of Ta$_{1-y}$Pd$_y$ has never been studied before, the Ta-H phase diagram is well-known for bulk Ta \cite{pryde1971,sanmartin1991,schober1996}. Below the critical temperature of $T_C$ = 61~$\degree$C various ordered and disordered cubic and orthorhombic phases are found in bulk TaH$_x$ for various values of $x$, and the transitions between these phases are known to be of first-order and involve substantial hysteresis.

For the nanosized thin films, the sensing measurements reveal a hysteresis free response over the entire pressure window at room temperature, suggesting that the various phase transitions are suppressed. The \textit{in-situ} XRD measurements of Figure \ref{InSituXRD} confirm this: with increasing hydrogen pressure, we find a continuous shift of the diffraction peak to lower angles (Figure \ref{InSituXRD}(a)), indicating a gradual hydrogenation and expansion of the TaH$_x$ bcc unit cell. In addition, and strikingly different from bulk Ta, we do not find any sign of hysteresis when we stepwise decreased the hydrogen pressure. The same trends are found for Ta$_{0.88}$Pd$_{0.12}$ (Figure \ref{InSituXRD}(b,c)), showing that for both materials no structural phase transition occurs upon exposure to hydrogen. As such, these diffraction results indicate a wide hydrogen solubility window over a large pressure range within one single thermodynamic phase. Based on the linear scaling between the changes in optical transmission $\ln(\mathcal{T}/\mathcal{T}_{prep})$ and the metal-to-hydrogen ratio $x$ as derived through \textit{in-situ} neutron reflectometry experiments \cite{bannenberg2019,bannenberg2020}, we find that TaH$_x$ hydrogenates at room temperature and at $P_{H2}$ = 10$^6$~Pa to $x$ $\approx$ 0.8. This implies a solubility window of $\Delta$x $\approx$ 0.8 that is substantially larger than for instance thin film Pd$_{1-y}$Au$_y$ or Hf. Indeed, these two most competitive metal hydride hydrogen sensing materials both have a solubility window of $\Delta$x $\approx$ 0.4. The combination of a wide solubility window and not signs of any hysteresis is highly beneficial for hydrogen sensing: it allows for the hysteresis-free sensing of hydrogen in a wide pressures window of over 7 orders in magnitude with relatively short response times that are not hindered by the nucleation of domains.

The suppression of the first-order phase transitions in sputtered Ta thin films highlights the profound impact of nanostructering on the properties of metal hydrides. Accommodating hydrogen in the Ta lattice results in lattice expansion, inducing strain in the host metal lattice. Different from bulk materials, two-dimensional clamped films have to obey constraints on lateral expansion as expansion can only be realized in the out-of-plane direction, which may result in a very high in-plane stress \cite{pundt2006,pivak2009,wagner2008,wagner2016,burlaka2016}. In addition, the nucleation of domains, inducing locally large stresses, may also be hindered considerably by the clamping of the film to the substrate \cite{pundt2006,mooij2013,bannenberg2016,baldi2018,bannenberg2019PdAu,burlaka2016,wagner2019}, with the ultimate result that the critical temperature is effectively reduced. Remarkable is that the reduction of the critical temperature is not accompanied by the introduction of hysteresis arising from plastic deformation as is e.g. the case in Pd and Pd$_{1-y}$Au$_y$ thin films. This may either imply that the critical stress to suppress the phase transition in the highly-textured films is relatively low, or that the stress to induce plastic deformation is sufficiently high for Ta$_{1-y}$Pd$_y$. Irrespective of the scenario, the absence of hysteresis of any sort positions Ta$_{1-y}$Pd$_y$ as an attractive material for thin film hydrogen sensing applications.

The fact that nanostructuring suppresses the undesired first order phase transitions in Ta may imply that hysteresis-free hydrogen sensing is not possible in other nanoconfinements as e.g. in the form of Ta nanoparticles deposited on a substrate  or when the thickness of the film is altered. Indeed, previous research on metal hydrides has shown that the way a metal hydride is nanostructured strongly affects the hydrogen solubility and the presence of phase transitions \cite{feenstra1983,pivak2009,bloch2010,baldi2014,syrenova2015,griessen2016,burlaka2016,wagner2016,bannenberg2019PdAu}. However, the alloying of tantalum with Pd also results in a smaller unit cell, with the result that the critical temperature is likely to be lowered as a result of this compression. As such, this would make hysteresis-free hydrogen sensing with Ta$_{1-y}$Pd$_y$ possible with other geometries than thin films such as nanoparticles in combination with a frequency-modulated localized surface plasmon resonance optical readout.

%

\section{Conclusion}
In conclusion, Ta$_{1-y}$Pd$_y$ is an effective and versatile hydrogen sensing material that has a sensing range of at least seven orders of magnitude in pressure both at room and elevated temperatures. Nanoconfinement of the Ta$_{1-y}$Pd$_y$ layer suppresses the series of first-order phase transitions present in bulk and ensures a sensing response free of any hysteresis within a single thermodynamic phase. The alloying with Pd compresses the unit cell and is effective in tuning the sensing range without a loss of sensitivity of the sensor. In combination with suitable and rationally designed capping layers, it features sub-second response times at room temperature that are faster than any reported thin film sensing material. The combination of these short response times, large sensing range and possibility for cost-effective production paves the way for a large-scale implementation of this material in a sustainable hydrogen powered economy. 

\section{Experimental Section}
\subsection{Sample preparation}
The Ta$_{1-y}$Pd$_y$ thin film samples are composed of a 4~nm titanium adhesion layer, a 40~nm Ta$_{1-y}$Pd$_y$ sensing layer and a 10~nm capping layer to catalyze the hydrogen dissociation and recombination reaction and prevent the film from oxidation. As a capping layer, we  used either a (i) single 10~nm Pd layer or a (ii) 10~nm Pd$_{0.6}$Au$_{0.35}$Cu$_{0.05}$ layer covered with a 30~nm PTFE layer to reduce response times\cite{ngene2014}. The layers are deposited on 10 $\times$ 10~mm$^2$ quartz substrates (thickness of 0.5~mm and surface roughness $<$ 0.4~nm) in 0.3~Pa of Ar by magnetron sputtering in an ultrahigh vacuum chamber (AJA Int.) with a base pressure of 10$^{-10}$~Pa. The substrates are rotated to enhance the homogeneity of the deposited layers. Typical deposition rates include 0.13~nm s$^{-1}$ (50~W DC) for Pd, 0.10~nm s$^{-1}$ (125~W DC) for Ta, 0.05~nm s$^{-1}$ (100~W DC) for Ti, 0.11~nm s$^{-1}$ (25~W DC) for Au and 0.08~nm s$^{-1}$ (25~W DC) for Cu. The deposition rates are determined by sputtering each target independently at a fixed power over a well-defined time interval. Subsequently, X-ray reflectometry (XRR) was used to determine the layer thickness of this reference sample, from which the sputter rate was computed. Differently, PTFE was deposited by radiofrequency magnetron sputtering in 0.5~Pa of Ar and the thickness of the reference film was measured with a DekTak3 profilometer. The Ta target was pre-sputtered for at least 240~min to avoid possible contamination from the tantalum oxide and nitride layers present at the surface of the target. We note that for large-scale manufacturing one can use alloy targets, and that alloys of Ta$_{1-y}$Pd$_y$ are, even at low concentrations of Pd, not susceptible to nitration and oxidation. This is one of the advantages of the Ta$_{1-y}$Pd$_y$ over single-element Ta.

The quality and thickness of all samples was verified with XRD and XRR (Figure \ref{XRR}). The fits to the XRR data (see below for experimental details) reveal that the deviation of the layer thickness between the different samples is at maximum 3\%, that the density of the various layers is consistent with the literature value for bulk material, and that the roughness of the surface of the capping layer is at maximum 1.5~nm. 

Prior to the measurements, we exposed the thin films to three cycles of hydrogen with a maximum pressure of $P_{H2}$  = 10$^{+6}$~Pa at $T$ = 28~$\degree$C. Reproducible and hysteresis-free results are obtained from the second cycle onwards. Differences between the first and subsequent cycles are common to thin film metal hydrides, as in general, a few cycles of exposure to hydrogen are required to show reproducible results due to a settling of the microstructure. Indeed, we find that the d-spacing of the Ta$_{1-y}$Pd$_y$ layer decreases (Figure \ref{Dspacing_before_after}) and the preferred orientation improves (Figure \ref{Rocking_before_after_H2}) after exposure to hydrogen. 

\subsection{Optical measurements}
The white-light optical transmission of the Pd-capped samples were measured using hydrogenography \cite{gremaud2007hydrogenography} with a Sony DXC-390P three charge-coupled device (3CCD) color video camera and a maximum acquisition frequency of 0.5~Hz. The transmission is averaged over an area of 80~mm$^2$. Five Philips MR16 MASTER LEDs (10/50~W) with a color temperature of 4,000~K are used as a light source (Figure \ref{SpectrumLED}). The measurements on the samples capped with 10~nm Pd$_{0.6}$Au$_{0.35}$Cu$_{0.05}$ and 30~nm PTFE were performed using a similar set-up in which the 3CCD camera was replaced by an Imaging Source 1/2.5" Aptina CMOS 2592 $\times$ 1944 pixel monochrome camera with an Edmunds Optics 55-906 lens, i.e. the same camera as used in ref. \cite{bannenberg2016}, to achieve higher acquisition frequencies of up to 20~Hz. The transmission is averaged over an area of 180 $\times$ 180 pixels, corresponding to about 80~mm$^2$. A reference sample is used to compensate for fluctuations of the LED white light source. The partial hydrogen pressures of 10$^{-1}$ $<$ $P_{H2}$ $<$ 10$^{+6}$~Pa are obtained by using 0.1\%, 4\% and 100\% H$_2$ in Ar gas mixtures. Typical gas flows are 20~s.c.c.m. for increasing pressure steps and 100~s.c.c.m. for decreasing pressure steps. The wavelength dependent optical transmission was measured using an Ocean Optics HL-2000-FHSA halogen light source and an Ocean Optics Maya 2000 Pro spectrometer with an acquisition frequency of 2.5~Hz. 

\subsection{Structural measurements}
Ex-situ X-ray Reflectometry (XRR) and X-ray diffraction (XRD) measurements were performed with a Bruker D8 Discover (Cu-K$\alpha$ $\lambda$ = 0.1542~nm). The XRR measurements were fitted with GenX3 \cite{bjorck2007} to obtain estimates for the layer thickness, roughness and density of the thin films. \textit{In-situ} XRD measurements were performed with a Bruker D8 Advance (Co-K$\alpha$ $\lambda$ = 0.1789~nm) in combination with an Anton Paar XRK 900 reactor chamber. During these experiments, a mixture of 4\% H$_2$ in helium was used as a gas and a constant flow of at least 10~s.c.c.m. was maintained at all times.


\section*{Acknowledgments}
We thank Christiaan Boelsma for creating the TOC image and for fruitful discussions.

\bibliography{Hydrogen_Sensing}

\section{Supplementary Figures}
\renewcommand{\thefigure}{S\arabic{figure}}
\setcounter{figure}{0}
\begin{figure}[b]
\begin{center}
\includegraphics[width= 0.45\textwidth]{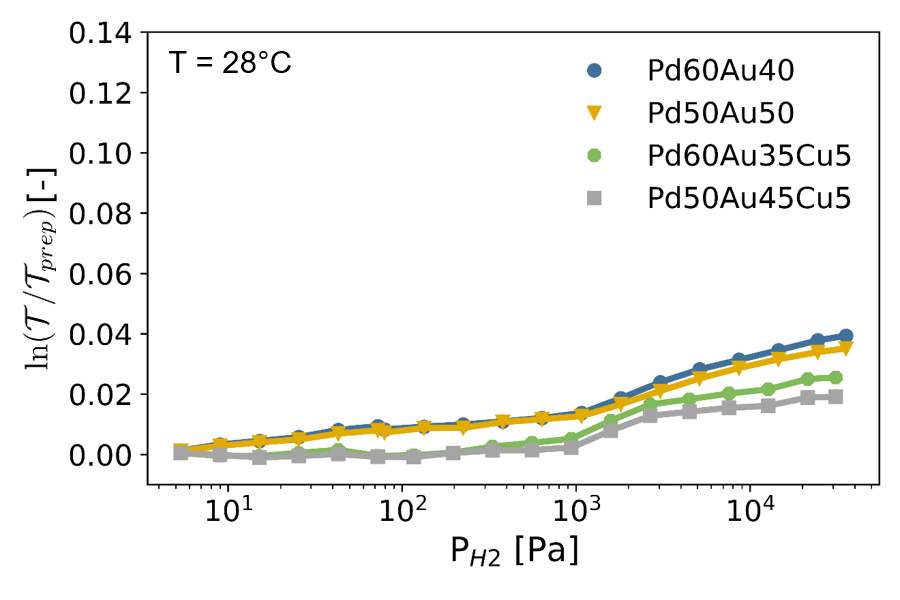}
\caption{Partial hydrogen pressure dependence of the optical responses of 10~nm Pd$_{1-y}$Au$_{y}$ and Pd$_{1-y-0.05}$Au$_{y}$Cu$_{0.05}$ capping layers at $T$ = 28~$\degree$C. The response was measured by stepwise increasing the pressure between $P_{H2}$ = 1 $\times$ 10$^{1}$ and 4 $\times$ 10$^5$~Pa. The optical response is measured as the natural logarithm of the changes of the white light optical transmission $\mathcal{T}$ relative to the transmission of the as-prepared film ($\mathcal{T}_{prep}$) and has been computed by subtracting the response of the 4~nm Ti adhesion layer and 40~nm Ta layer from the Pd$_{1-y}$Au$_{y}$ capped 40~nm Ta thin films with a 4~nm Ti adhesion layer (see Figure \ref{stepresponse}). Previous neutron reflectometry measurements indicate that $\ln$($\mathcal{T}/\mathcal{T}_{prep})$ is proportional to the the hydrogen content $x$ in the Pd$_{1-y}$Au$_{y}$H$_x$ layer \cite{bannenberg2019PdAu}.}
\label{OpticalResponseCap}
\end{center}
\end{figure}

\begin{figure*}[tb]
\begin{center}
\includegraphics[width= 0.6\textwidth]{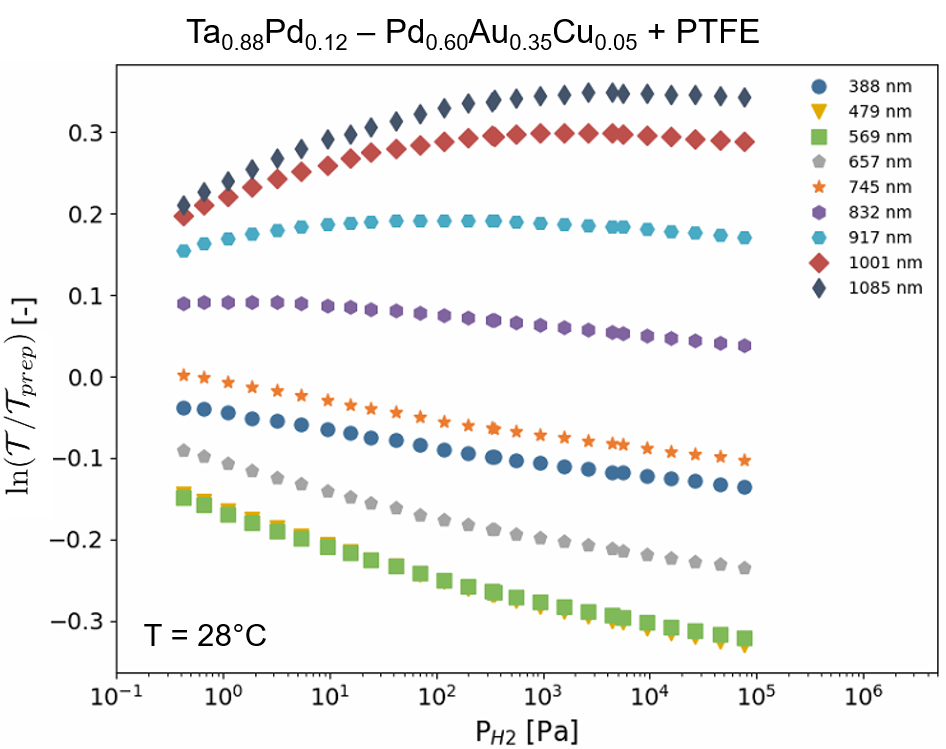}
\caption{Partial hydrogen pressure dependence of the optical transmission $\mathcal{T}$ measured at different wavelengths of a 40~nm Ta$_{0.88}$Pd$_{0.12}$ thin film with a 4~nm Ti adhesion layer capped with a 10~nm Pd$_{0.6}$Au$_{0.35}$Cu$_{0.05}$ layer that is covered with a 30~nm PTFE layer (`Ta$_{0.88}$Pd$_{0.12}$  - Pd$_{0.6}$Au$_{0.35}$Cu$_{0.05}$ + PTFE'). The changes in optical transmission are measured relative to the optical transmission of the as-prepared state ($\mathcal{T}_{prep}$).}
\label{WavelengthDependence}
\end{center}
\end{figure*}

\begin{figure}[tb]
\begin{center}
\includegraphics[width= 0.45\textwidth]{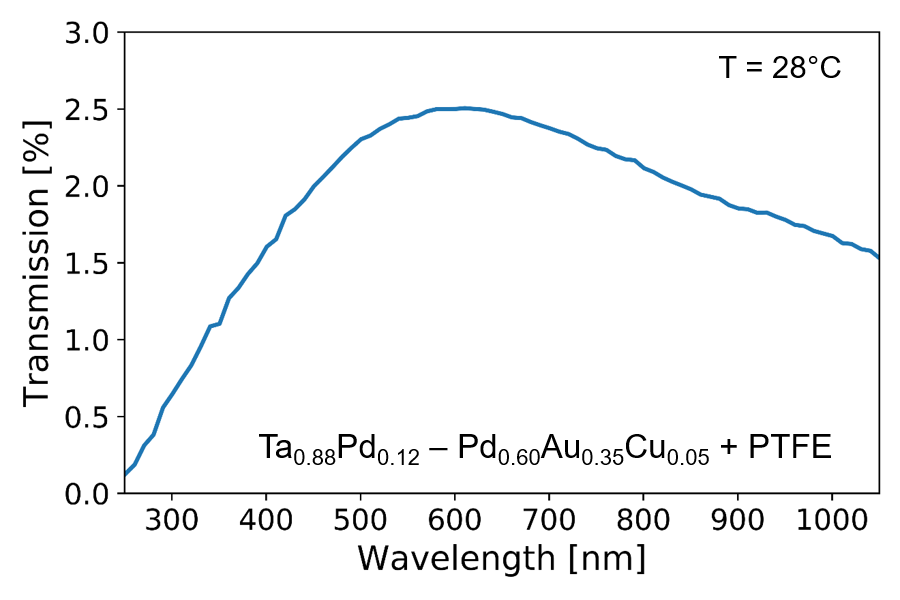}
\caption{Optical transmission of the as-prepared 40~nm Ta$_{0.88}$Pd$_{0.12}$ thin film with a 4~nm Ti adhesion layer and capped with a 10~nm Pd$_{0.6}$Au$_{0.35}$Cu$_{0.05}$ layer that is covered with a 30~nm PTFE layer a function of time.}
\label{Transmission}
\end{center}
\end{figure}

\begin{figure*}[tb]
\begin{center}
\includegraphics[width= 1\textwidth]{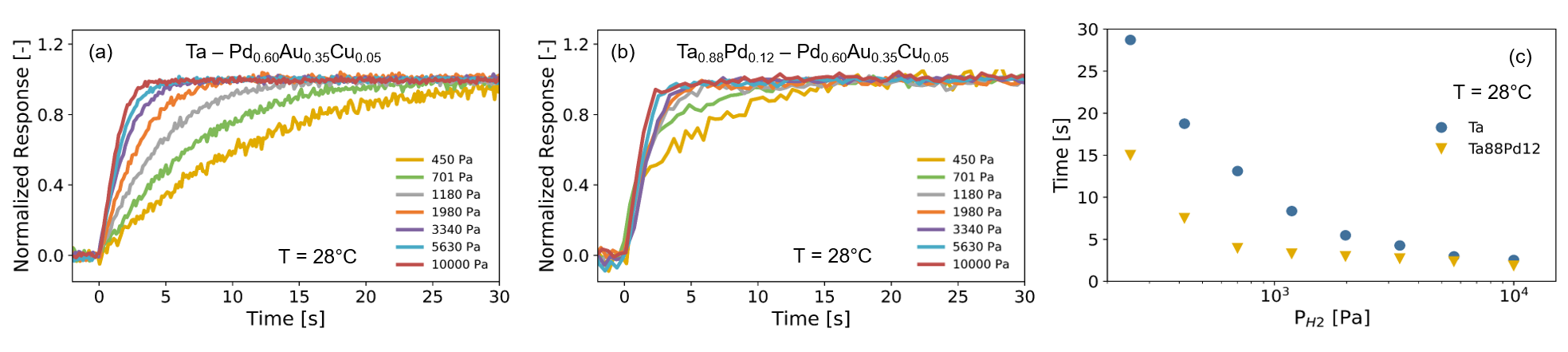}
\caption{Comparison of the absorption kinetics of a 40~nm Ta and a 40~nm Ta$_{0.88}$Pd$_{0.12}$ thin film that both have a 4~nm Ti adhesion layer and are both capped with a 10~nm Pd$_{0.6}$Au$_{0.35}$Cu$_{0.05}$ layer at $T$ = 28~$\degree$C. (a) Normalized responses of the 40~nm Ta thin film to a series of pressure steps between $P_{H2}$ = 0.5 10$^2$~Pa and the partial hydrogen pressure indicated. (a) Normalized responses of the 40~nm Ta$_{0.88}$Pd$_{0.12}$ thin film to a series of pressure steps between $P_{H2}$ = 0.5 10$^2$~Pa and the partial hydrogen pressure indicated. (c) Hydrogen pressure dependence of the response time of the thin films. The response time is defined as the time to reach 90\% of the total signal.}
\label{ResponseTimeComparison}
\end{center}
\end{figure*}

\begin{figure*}[tb]
\begin{center}
\includegraphics[width= 0.95\textwidth]{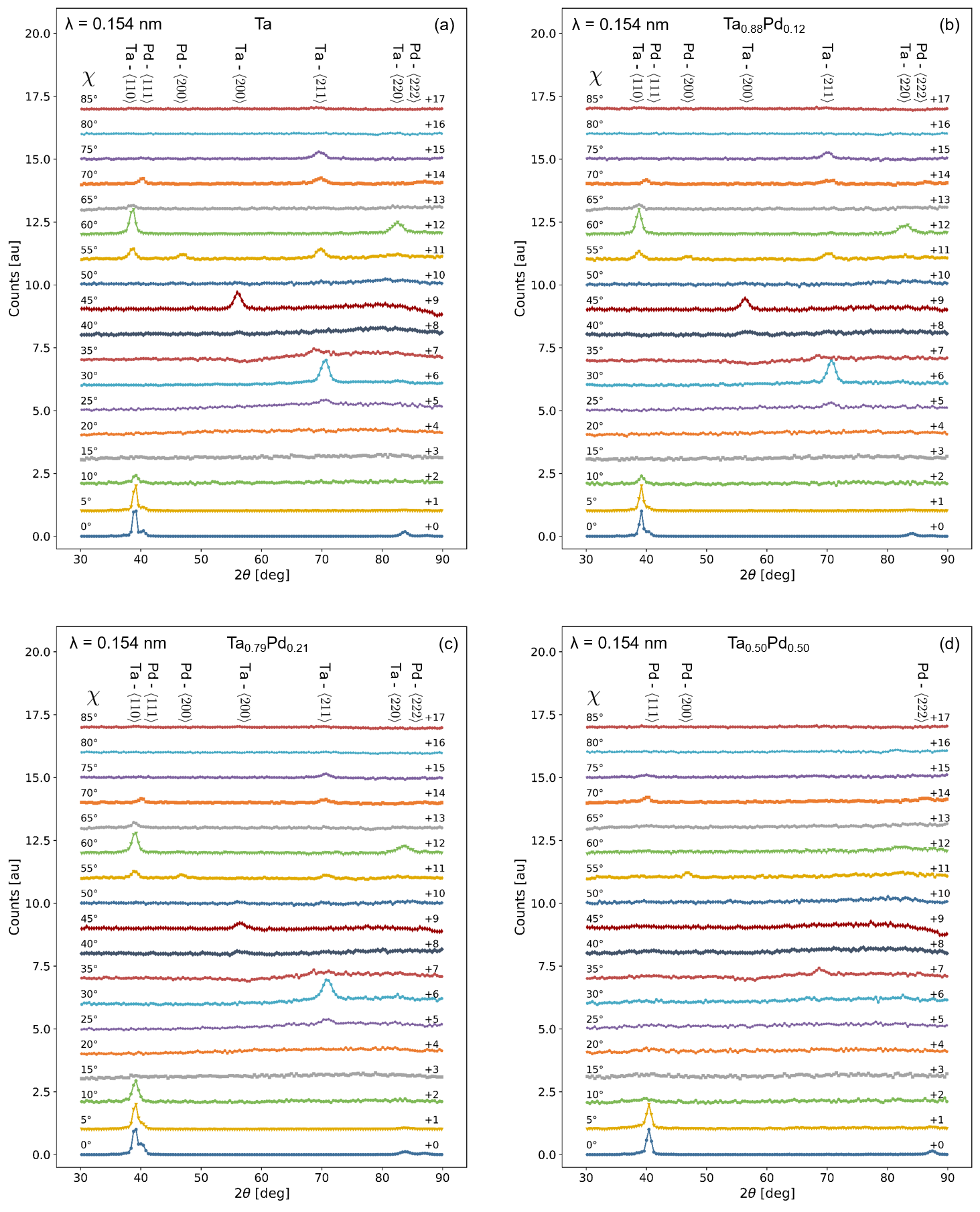}
\caption{Ex-Situ X-ray diffractograms (Cu-K$\alpha$ $\lambda$ = 0.1542~nm) at $T$ = 28~$\degree$C of (a) Ta, (b) Ta$_{0.88}$Pd$_{0.12}$, (c) Ta$_{0.79}$Pd$_{0.21}$ and (d) Ta$_{0.50}$Pd$_{0.50}$ after exposure to at least 200 cycles of hydrogen and measured when the sample was tilted perpendicular to the X-ray beam by different angles $\chi$. The X-ray diffractograms have been normalized to the larger of 0.4~cps or the maximum intensity and shifted by the values indicated.}
\label{ChiXRD}
\end{center}
\end{figure*}

%

\begin{figure}[tb]
\begin{center}
\includegraphics[width= 0.45\textwidth]{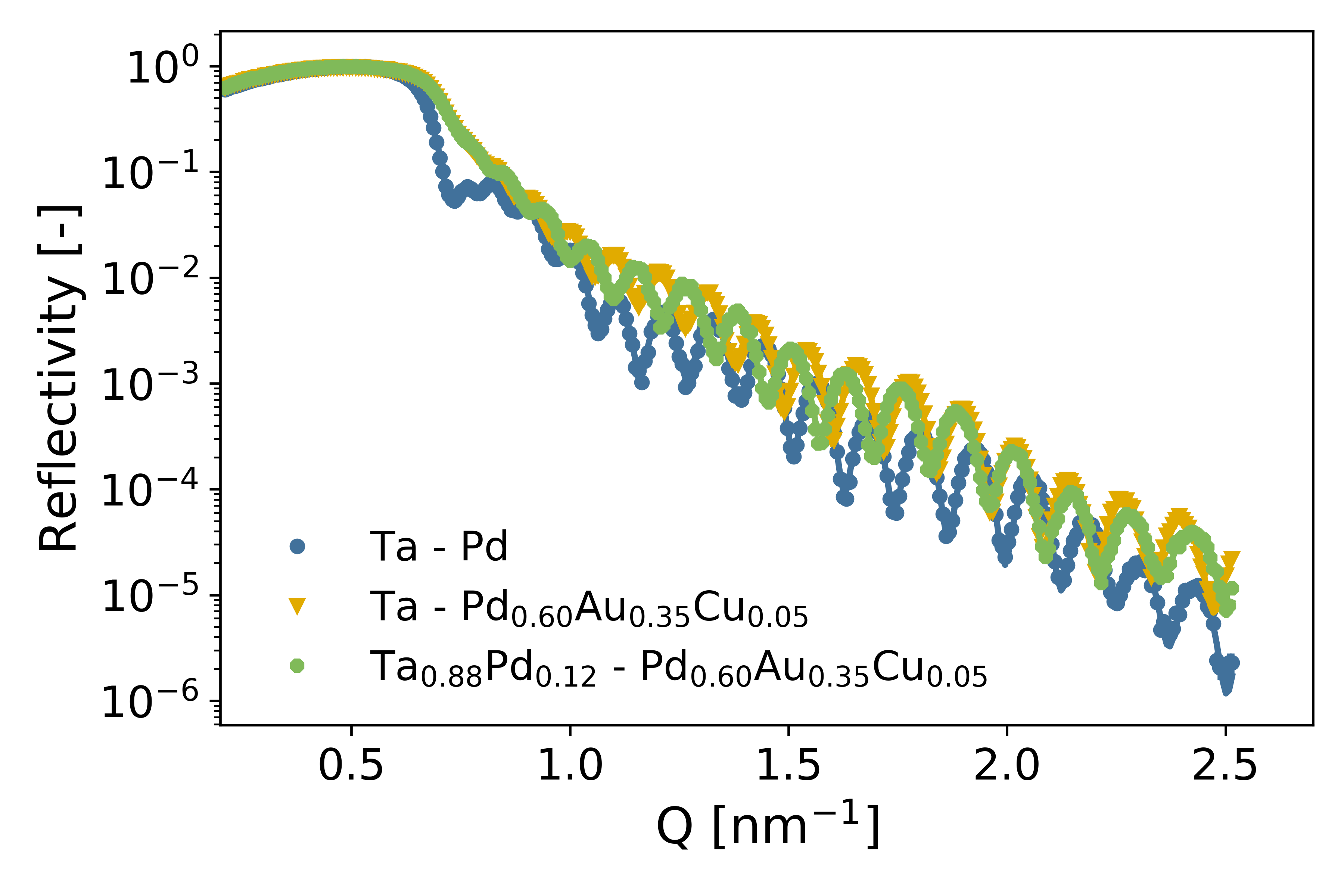}
\caption{X-Ray Reflectometry (XRR) measurements of the as-prepared Ta$_{1-y}$Pd$_y$ thin films with a 4~nm Ti adhesion layer and for the compositions and capping layers indicated in the figure legend. The solid lines indicate fits to the measured data. }
\label{XRR}
\end{center}
\end{figure}

\begin{figure}[tb]
\begin{center}
\includegraphics[width= 0.45\textwidth]{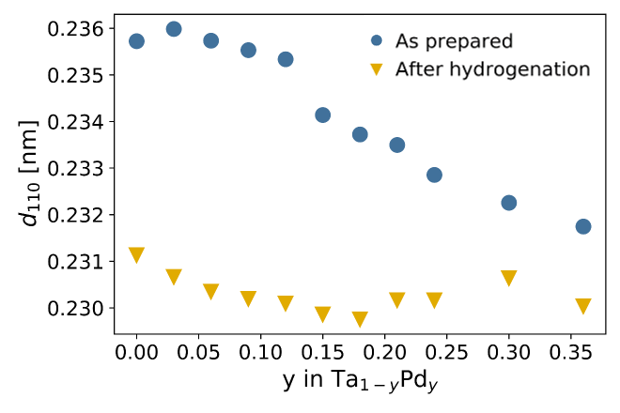}
\caption{Pd doping dependence of the d$_{110}$-spacing in Ta$_{1-y}$Pd$_{y}$ before and after exposure to hydrogen.}
\label{Dspacing_before_after}
\end{center}
\end{figure}

\begin{figure*}[tb]
\begin{center}
\includegraphics[width= 0.85\textwidth]{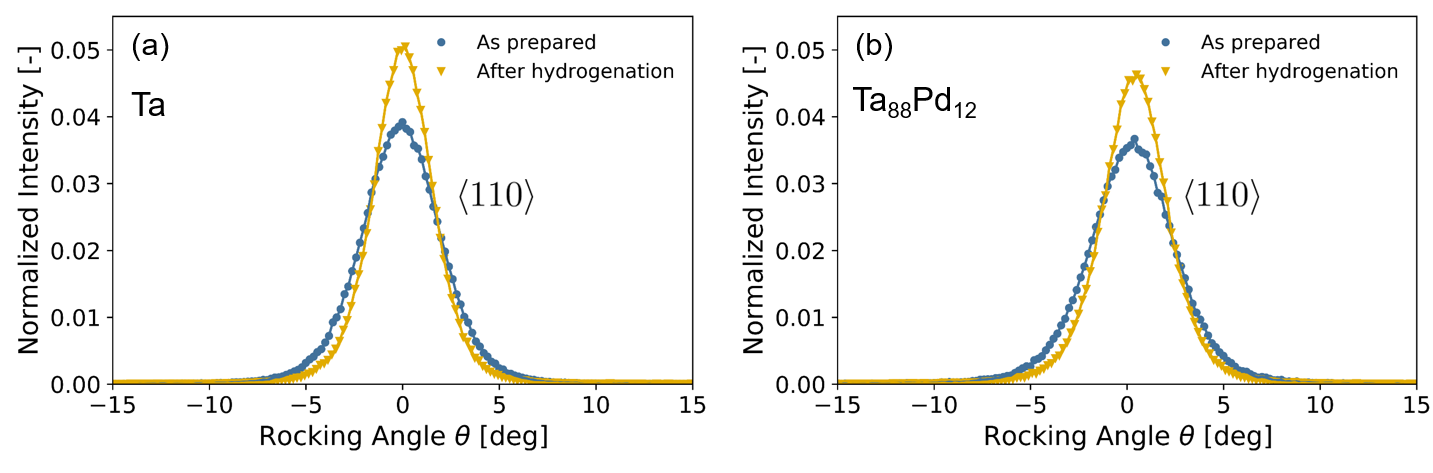}
\caption{Rocking curves around the $\langle$110$\rangle$ diffraction peak of Pd-capped (a) and (b) Ta$_{0.88}$Pd$_{0.12}$ thin films with a 4~nm Ti adhesion layer in the as-prepared state and after exposure
to at least 200 cycles of hydrogen at $T$ = 28~$\degree$C.}
\label{Rocking_before_after_H2}
\end{center}
\end{figure*}

\begin{figure}[tb]
\begin{center}
\includegraphics[width= 0.45\textwidth]{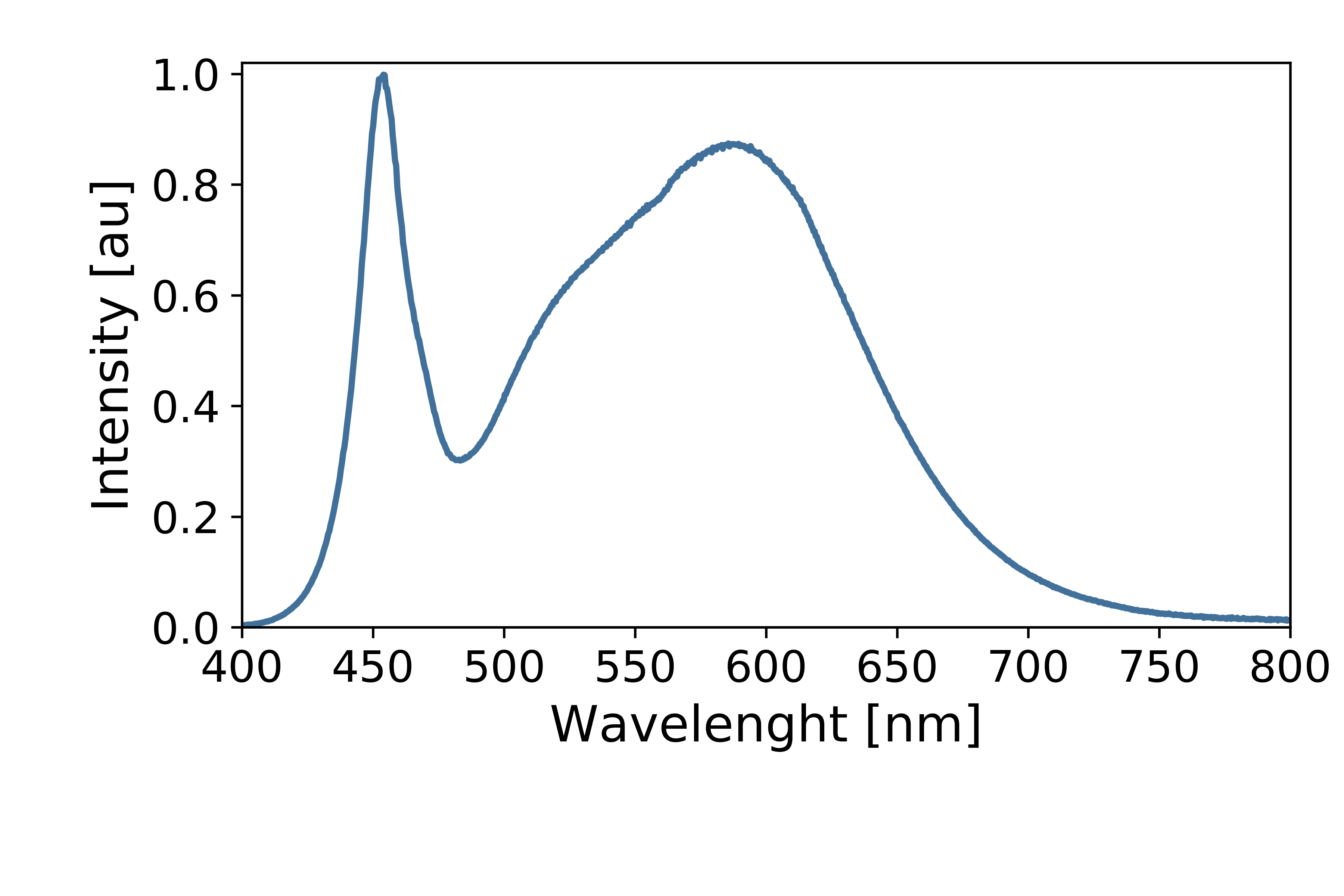}
\caption{Spectrum of the Philips MR16 MASTER LEDs (10/50~W) with a color temperature of 4,000~K used for the white-light optical transmission (hydrogenography) measurements.}
\label{SpectrumLED}
\end{center}
\end{figure}

\end{document}